\documentclass[sigconf]{acmart}

\newtheorem{theorem}{Theorem}
\usepackage{enumitem}
\usepackage{algorithm}
\usepackage{algpseudocode}

\usepackage{colortbl}
\usepackage{multirow}
\usepackage{graphicx}
\usepackage[normalem]{ulem}
\useunder{\uline}{\ul}{}

\definecolor{gblue}{RGB}{66,133,244}

\newcommand{\emphcolorbluee}[1]{\colorbox{gblue!10}{#1}}

\newcommand{\emphcolorbrownn}[1]{\colorbox{brown!15}{#1}}

\newcommand{\emphcolorgreenn}[1]{\colorbox{green!8}{#1}}

\usepackage{arydshln}
\makeatletter
\def\adl@drawiv#1#2#3{%
        \hskip.5\tabcolsep
        \xleaders#3{#2.5\@tempdimb #1{1}#2.5\@tempdimb}%
                #2\z@ plus1fil minus1fil\relax
        \hskip.5\tabcolsep}
\newcommand{\cdashlinelr}[1]{%
  \noalign{\vskip\aboverulesep
           \global\let\@dashdrawstore\adl@draw
           \global\let\adl@draw\adl@drawiv}
  \cdashline{#1}
  \noalign{\global\let\adl@draw\@dashdrawstore
           \vskip\belowrulesep}}
\makeatother

\usepackage[normalem]{ulem} 

\usepackage{colortbl} 
\definecolor{mygray}{rgb}{0.9, 0.9, 0.9}
\definecolor{myred}{rgb}{0.68627451, 0.14117647, 0.09803922}
\newcommand{\myeq}[1]{{Eq.~(\ref*{eq:#1})}}
\newcommand{\mysec}[1]{{Section~\ref*{sec:#1}}}
\newcommand{\mytab}[1]{{Table~\ref*{tab:#1}}}
\newcommand{\myfig}[1]{{Fig.~\ref*{fig:#1}}}
\newcommand{\myappendix}[1]{{Appendix~\ref*{appendix:#1}}}

\newcommand{\mytheorem}[1]{{Theorem~\ref*{theorem:#1}}}

\AtBeginDocument{%
  }

\setcopyright{acmlicensed}
\copyrightyear{2025}
\acmYear{2025}
\setcopyright{acmlicensed}\acmConference[WWW '25]{Proceedings of the ACM Web Conference 2025}{April 28-May 2, 2025}{Sydney, NSW, Australia}
\acmBooktitle{Proceedings of the ACM Web Conference 2025 (WWW '25), April 28-May 2, 2025, Sydney, NSW, Australia}
\acmDOI{10.1145/3696410.3714524}
\acmISBN{979-8-4007-1274-6/25/04}

\begin{document}

\title{SPRec: Self-Play to Debias LLM-based Recommendation}

\author{Chongming Gao}
\authornote{Both authors contributed equally to this research.}
\email{chongminggao@ustc.edu.cn}
\affiliation{%
  \institution{University of Science and Technology of China}
  \city{Hefei}
  \country{China}
}
\orcid{0000-0002-5187-9196}

\author{Ruijun Chen}
\authornotemark[1]
\orcid{0009-0009-0186-9561}
\email{rjchen20@mail.ustc.edu.cn}
\affiliation{%
  \institution{University of Science and Technology of China}
  \city{Hefei}
  \country{China}
}

\author{Shuai Yuan}
\email{syuanaf@connect.ust.hk}
\orcid{0000-0001-6730-5755}
\affiliation{%
  \institution{Hong Kong University of Science and Technology}
  \city{Hong Kong}
  \country{China}
}

\author{Kexin Huang}
\email{huangkx@mail.ustc.edu.cn}
\orcid{0009-0001-4868-0952}
\affiliation{%
  \institution{University of Science and Technology of China}
  \city{Hefei}
  \country{China}
}

\author{Yuanqing Yu}
\email{yyq23@mails.tsinghua.edu.cn}
\orcid{0000-0002-2942-1576}
\affiliation{%
  \institution{Tsinghua University}
  \city{Beijing}
  \country{China}
}

\author{Xiangnan He}
\authornote{Corresponding Author.}
\email{xiangnanhe@gmail.com}
\orcid{0000-0001-8472-7992}
\affiliation{%
  \institution{MoE Key Lab of BIPC, University of Science and Technology of China}
  \city{Hefei}
  \country{China}
}

\renewcommand{\shortauthors}{Gao and Chen et al.}

\begin{abstract}

Large language models (LLMs) have attracted significant attention in recommendation systems. Current work primarily applies supervised fine-tuning (SFT) to adapt the model for recommendation tasks. However, SFT on positive examples only limits the model's ability to align with user preference. To address this, researchers recently introduced Direct Preference Optimization (DPO), which explicitly aligns LLMs with user preferences using offline preference ranking data. However, we found that DPO inherently biases the model towards a few items, exacerbating the filter bubble issue and ultimately degrading user experience.

In this paper, we propose SPRec, a novel self-play framework designed to mitigate over-recommendation and improve fairness without requiring additional data or manual intervention. In each self-play iteration, the model undergoes an SFT step followed by a DPO step, treating offline interaction data as positive samples and the predicted outputs from the previous iteration as negative samples. This effectively re-weights the DPO loss function using the model's logits, adaptively suppressing biased items. Extensive experiments on multiple real-world datasets demonstrate SPRec's effectiveness in enhancing recommendation accuracy and fairness.
The code is available via \url{https://github.com/RegionCh/SPRec}.

\end{abstract}




\maketitle

\section{Introduction}
Recently, large language models (LLMs) have demonstrated significant success across numerous domains, showcasing advanced capabilities in learning, reasoning, and generalizing to downstream tasks \cite{wang2024towards,kddbias2024}. In the field of recommender systems, there has been growing interest in leveraging the potential of LLMs \cite{surveyLLM4rec}. One prominent approach involves positioning LLMs as the central recommendation backbone, utilizing users' past interactions and current needs to generate personalized recommendations \cite{bao2023bi,LLMRec}. Compared to traditional methods, LLM-based recommendation systems (LRSs) offer distinct advantages, including a deeper contextual understanding and the flexibility to adapt to users’ evolving preferences.

\begin{figure}[!t]
  \centering
  \includegraphics[width=0.95\linewidth]{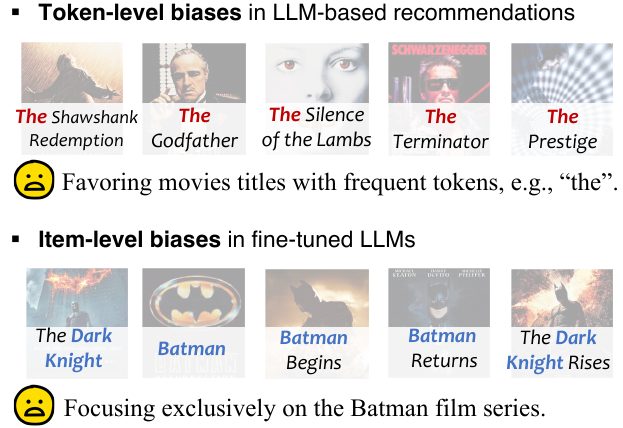}
  \caption{Homogeneity issues in LLM-based recommendation results caused by token-level and item-level biases.}
  \vspace{-2mm}
  \label{fig:intro}
\end{figure}

To enable LLMs to learn collaborative filtering signals and effectively perform item recommendations, a prevalent strategy is to fine-tune pre-trained LLMs via Supervised Fine-Tuning (SFT) \cite{bao2023bi}. This approach allows LLMs to efficiently internalize user preferences from offline data by adjusting their parameters to align with the recommendation task. Building on SFT, recent research has adopted Direct Preference Optimization (DPO) \cite{rafailov2024direct} to further refine user preferences \cite{chen2024softmax, DMPO, liao2024rosepo}. While SFT relies solely on desirable answers, DPO incorporates both chosen and rejected response pairs, allowing the LLM to learn user ranking preferences and gain a more nuanced understanding of fine-grained, personalized information. This approach mirrors the common practice in recommendation models, which utilize both positive and negative samples for effective training \cite{10.1145/3522672,shi2023negative}.

Despite these advancements, we find that employing DPO to align user preferences in recommender systems inherently introduces significant biases due to its underlying mechanisms.
These biases can lead to serious homogeneity issues, where LLMs recommend items with similar names or content. \myfig{intro} illustrates how token-level and item-level biases manifest in the Top-$K$ movie recommendation. 
Token-level biases arise due to the fact that LLMs generate item names in a tokenized fashion. Since LLMs are usually tuned to maximize the likelihood of target tokens, items with more common tokens (e.g., movies with the word ``the'' in their titles) may be overrepresented, regardless of user relevance.
At the item level, biases can emerge from multiple factors, particularly after fine-tuning, where LLMs may disproportionately recommend popular items, such as the Batman film series. This can lead to filter bubbles \cite{gao2023alleviating,gao2023cirs}, where users are repeatedly exposed to a narrow range of popular content, limiting the diversity of recommendations and degrading the user experience.

Some research has been conducted on bias and unfairness issues in LRSs \cite{kddbias2024,jizhi2023fair,gallegos-etal-2024-bias}. \citet{kddbias2024} provided a comprehensive overview of the various types of biases that emerge across different stages of these models and outlined strategies to mitigate them. For example, \citet{jiang2024item} proposed re-weighting the fine-tuning loss for each item and re-ranking the generated results to ensure equitable treatment across genre groups. Similarly, \citet{bao2024decoding} adjusted the LLM decoding process by removing the length normalization term and incorporating predictions from a text-free model, helping to reduce amplification bias and address homogeneity issues. However, these methods often rely on carefully crafted rules or external knowledge, limiting their broader applicability in general recommendation systems.

To this end, we propose a self-play recommendation tuning framework, SPRec, to adaptively suppress biases and improve fairness in LRSs without the need for additional data or expert knowledge. The core idea of SPRec is straightforward: each tuning iteration begins with an SFT round using positive samples from offline data, followed by a DPO. In the DPO step, the SFT data is treated as positive samples, while the predicted outputs from the previous iteration are treated as negative samples. The philosophy is to let the model ``play'' with its own output by re-weighting the DPO loss function based on its predictions. As a result, items that rank higher in the model's predictions are penalized, while the SFT process reinforces the ranking of positive items. Over time, this self-play learning process adaptively suppresses undesirable items (biases) while maintaining alignment with positive samples. Extensive experiments on public datasets demonstrate that SPRec effectively improves both accuracy and fairness, showcasing its potential as a practical and efficient solution for LRSs.

The main contributions of this paper are as follows:
\begin{itemize}
    \item We analyze how current LRSs tuned through DPO inevitably exhibit biases due to their underlying learning mechanisms, leading to the homogeneity issue.
    \item  We propose SPRec, a self-play recommendation tuning framework that addresses these biases and improves fairness without the need for external knowledge.
    \item Experiments validate SPRec improves accuracy, diversity, and fairness, with ablation studies indicating that the self-play negative samples contribute significantly to the improvements.
\end{itemize}

\section{Related work}
\label{appendix:related}
We provide a brief overview of LLM-based recommender systems and their associated bias issues, followed by an introduction to the self-play mechanism employed in our method.

\subsection{LLMs for Recommendation}

LLMs have shown exceptional generative, generalization, and reasoning capabilities in NLP, driving research into their applications for personalized recommendations. Their integration into recommendation tasks follows three main paradigms: (1) acting as decision makers \cite{bao2024decoding,RPP_TOIS}, (2) assisting by providing contextual information \cite{liu2024large,10.1145/3626772.3657974}, and (3) serving as user simulators \cite{10.1145/3589334.3645537,cai2024flow}. Early studies explored prompt engineering to leverage LLMs for recommendation tasks \cite{gao2023chat,zeroshot}.  

Later, fine-tuning methods emerged, demonstrating that adapting LLM parameters on recommendation data significantly enhances performance. These approaches primarily rely on SFT \cite{bao2024decoding,chen2025dlcrec}. To further align LLMs with user preferences, DPO has been employed for post-training \cite{chen2024softmax, DMPO, liao2024rosepo}. However, prior work has overlooked DPO's inherent tendency to introduce severe biases, favoring only frequently exposed items and degrading user experience. In this work, we are the first to identify this issue and propose a mitigation strategy.

\subsection{Biases in Recommender Systems}

Bias and fairness issues are pervasive in recommender systems and have been extensively studied. \citet{chen2023biases} provide a comprehensive survey on biases such as popularity bias, selection bias, and position bias. These biases can significantly impact user satisfaction, promoting clickbait content or reinforcing filter bubbles that reduce engagement \cite{gao2023cirs}. Additionally, algorithmic decisions may favor certain items, raising fairness concerns \cite{tois2023fairness,10.1145/3610302}, disproportionately affecting user groups and discouraging content creators \cite{yao2024unveiling,10.5555/3666122.3666764}.  

These challenges persist in LLM-based recommender systems \cite{surveyLLM4rec, xu2024study, FairnessMatters}. Research shows that LLMs can inherit social biases, leading to unfair recommendations related to sensitive attributes like gender and race \cite{jizhi2023fair}. \citet{kddbias2024} provide a unified distribution mismatch perspective on bias and fairness in information retrieval.  
Existing bias mitigation methods in LRSs typically rely on predefined target distributions or external guidance for LLM alignment. In contrast, we introduce the first self-play framework for mitigating bias in LRSs, requiring neither prior knowledge nor additional models. By simply modifying the tuning process, our approach reduces long-tail effects and improves fairness.




\subsection{Self-Play Mechanism}
Machine learning models are often data-driven, relying heavily on the availability of offline data. However, offline data is inherently limited, raising an important question: can algorithms improve themselves iteratively without the need for additional data? This is precisely the challenge that the self-play mechanism aims to address. The concept of self-play originated from board games such as Go and chess, exemplified by groundbreaking systems like AlphaGo Zero \cite{alphago_zero} and AlphaZero \cite{alphazero}.

In the era of LLMs, early preference alignment algorithms like RLHF and DPO operate as single optimization procedures. Building on this, \citet{chen2024self} proposed that LLMs can refine their capabilities through self-play by interacting with instances of themselves. More specifically, the LLM generates its own training data from prior iterations and subsequently learns a new policy that outperforms the old one \cite{SPPO}. This iterative fine-tuning framework has been shown to be equivalent to finding the Nash equilibrium of a two-player game, gaining significant recognition due to its solid theoretical foundation and simplicity \cite{calandriello2024human}.
In recommender systems, we leverage the self-play mechanism to adaptively reduce biased items in a simple and effective manner.

\section{Preliminary}
\label{sec:preliminary}
In this section, we provide a brief overview of the technologies for aligning LLMs with the recommendation task. We then introduce the idea of evaluating the biases and unfairness in LRSs.

\subsection{Supervised Fine-tuning (SFT)}
To enable an open-source LLM to learn recommendation tasks effectively, a practical approach is fine-tuning all or part of its parameters using demonstration data from offline recommendation logs. The objective is to align the model’s behavior with the recommendation task by maximizing the log-likelihood over the training dataset $\mathcal{D}$:
\begin{equation}
\label{eq:sft}
    \pi_{SFT} = \arg\max_{\pi_\theta} \mathbb{E}_{(x_i, y_i) \sim \mathcal{D}} \log\pi_\theta(y_i|x_i),
\end{equation}
where $(x_i, y_i)$ are input-output pairs from $\mathcal{D}$, with $x_i$ representing user context and interaction history, and $y_i$ the target item. Defining $p_\mathcal{D}(y|x)$ as the empirical probability (i.e., item popularity), SFT aligns model predictions by minimizing the forward KL-divergence:
\begin{align}
\label{eq:forward_KL}
    \pi_{SFT} &= \arg\min_{\pi_\theta} \mathbb{D}_{KL}(p_\mathcal{D}(y|x),\pi_\theta(y|x)) \notag  \\
    &= \arg\min_{\pi_\theta} \mathbb{E}_{(x_i,y_i)\sim \mathcal{D}} -\log\pi_\theta(y_i|x_i) + H(p_\mathcal{D}),
\end{align}
where $H(p_\mathcal{D})$ is the constant entropy of $p_\mathcal{D}$. 
\subsection{Direct Preference Optimization (DPO)}

To ensure that model outputs align with intricate user preferences, researchers have proposed Direct Preference Optimization (DPO) \cite{rafailov2024direct}, which optimizes the following objective function:
\begin{equation}
\label{eq:dpo}
    \min_{\pi_\theta} - \mathbb{E}_{(x, y_w, y_l) \sim \mathcal{D}} 
    \log \sigma \Bigg[
    \! \beta \log\!\left(\!\frac{\pi_\theta(y_w|x)}{\pi_\mathrm{ref}(y_w|x)}\!\right) 
    - \beta \log\!\left(\!\frac{\pi_\theta(y_l|x)}{\pi_\mathrm{ref}(y_l|x)}\!\right)\!
    \Bigg],
\end{equation}
where $(x, y_w, y_l)$ denotes a prompt $x$ with a chosen (preferred) answer $y_w$ and a rejected answer $y_l$. The parameter $\beta$ acts as a regularization factor, controlling the extent to which the learned policy $\pi_\theta$ deviates from the reference policy $\pi_\mathrm{ref}$.

In the context of recommendation tasks, $x$ represents the user context, typically comprising user features and historical interaction sequences, while $y_w$ and $y_l$ correspond to positive and negative samples, respectively.
The goal is to encourage the model to assign higher probabilities to preferred items ($y_w$) over less desirable ones ($y_l$), effectively capturing user preferences.

DPO offers an efficient and stable solution for preference alignment, eliminating the need for complex reward models often required in reinforcement learning-based approaches. Its natural ability to incorporate both positive and negative samples makes it particularly well-suited for recommender systems, where learning from contrasting user interactions is crucial \cite{chen2024softmax, DMPO, liao2024rosepo}.


\subsection{Evaluating Bias via Distribution Alignment}
When aligning user preferences, LLMs may inadvertently learn biased or unfair outcomes. To assess the bias and fairness issues in LRSs, a mainstream perspective is to formulate the problems as a mismatch distribution problem \cite{kddbias2024}. Specifically, let $R$ denote the ground-truth user preference (e.g., an item list), following the distribution $P(R)$, and let $\hat{R}$ represent the model-predicted preferences, drawn from $P(\hat{R})$. Bias or unfairness is then quantified by the mismatch between these two distributions:  $P(R) \neq P(\hat{R})$.

To apply this framework, we follow \citet{jiang2024item}, approximating $P(R)$ using the category distribution from offline training data. We further employ their MGU metric to systematically measure the degree of mismatch in our experiments.

\section{Problem of DPO: Amplify Popularity Bias}

We present an empirical analysis to demonstrate how DPO exacerbates popularity bias in LRSs, followed by a theoretical examination of the underlying mechanisms driving this phenomenon.

\begin{figure}[!t]
  \centering
   \includegraphics[width=\linewidth]{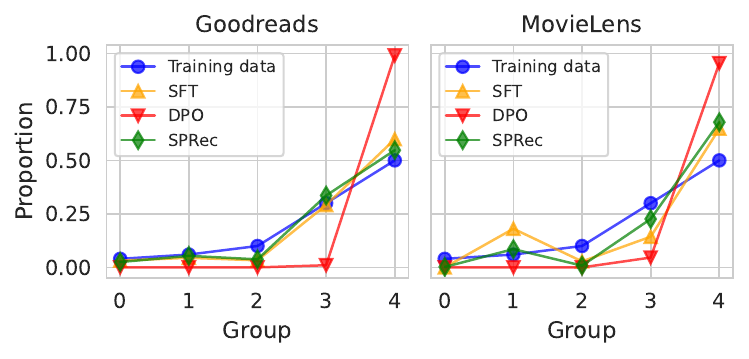}
  \caption{Distribution of cold-start recommendation results. Group 0: least popular, group 4: most popular.}
  \label{fig:Comparison}
\end{figure}

\begin{figure*}[t]
  \centering
  \includegraphics[width=0.92\linewidth]{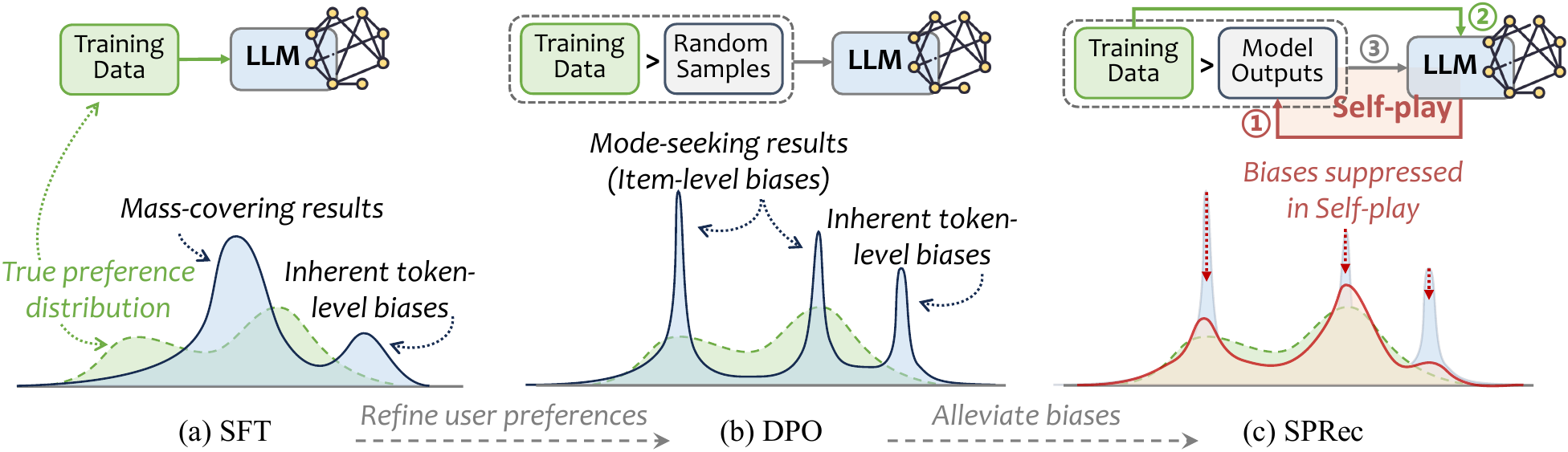}
  \caption{Illustration of SFT, DPO, and SPRec in LLM-based recommendations. (a) SFT generates mass-covering results but retains inherent biases. (b) DPO amplifies these biases by over-representing certain items. (c) SPRec mitigates bias through self-play, leveraging model outputs as negative samples to achieve balanced recommendations.}
\label{fig:method}
\end{figure*}

\subsection{Empirical Analysis}

To systematically evaluate how DPO amplifies recommendation bias, we design a cold-start recommendation task using two widely-used benchmark datasets: MovieLens and Goodreads\footnote{Dataset details are provided in \mysec{datasets}}. In this task, the LLM generates recommendations without access to user interaction history. We randomly sample 100 items and partition them into five groups based on interaction probabilities, ensuring balanced popularity levels. Positive samples are drawn accordingly, while negative samples for DPO training are randomly selected. The training and validation sets contain 4096 and 512 samples, respectively. After training, we analyze the distribution of recommendations across the five groups.

\myfig{Comparison} shows the proportion of recommendations allocated to each group before and after training, where group 0 represents the least popular items and group 4 the most popular. The results reveal three key insights: (1) SFT introduces a slight bias toward group 4; (2) DPO significantly amplifies this bias, causing recommendations to concentrate almost entirely on the most popular items; and (3) our proposed method, SPRec (introduced in detail later), effectively mitigates the bias amplification induced by DPO.

\subsection{Theoretical Analysis}
In recommendation tasks, the input and positive samples $(x, y_w)$ are derived from logged interactions in offline data, while negative samples $y_l$ are drawn from non-interacted items. Given an input $x$, the probability $p_\mathcal{D}(y|x)$ represents the conditional popularity of item $y$ in the dataset, and $q_\mathcal{D}(y|x)$ denotes the probability of the same item $y$ being selected as a negative sample. Using these definitions, the DPO loss in \myeq{dpo} can be rewritten as:
\begin{equation}
\label{eq:dpo_new}
\begin{gathered}
\mathcal L_\mathrm{DPO}(\pi_\theta;\pi_{ref}) = -\mathbb{E}_{(x,y_w)\sim \mathcal{D}, y_l\sim q_\mathcal{D}(\cdot|x)}\ell(\pi_\theta,\pi_\mathrm{ref},x,y_w,y_l),\\
\text{with }\ell(\cdot) = \log\sigma[\beta\log(\frac{\pi_\theta(y_w|x)}{\pi_{ref}(y_w|x)})-\beta\log(\frac{\pi_\theta(y_l|x)}{\pi_{ref}(y_l|x)})].
\end{gathered}
\end{equation}
In this setting, the optimal policy has a closed-form solution, as stated in the following theorem:
\begin{theorem}
\label{theorem:dpo_theorem}
The optimal policy $\pi_\theta^*(\cdot|x)$ for the DPO loss defined in \myeq{dpo_new} is given by:
$$
\pi_\theta^*(y|x) \propto \pi_\mathrm{ref}(y|x)\cdot \left(\frac{p_\mathcal{D}(y|x)}{q_\mathcal{D}(y|x)}\right)^{1/\beta}.
$$
\end{theorem}


The proof is deferred to \myappendix{DPO}. This result highlights that the optimal policy is proportional to the reference policy $\pi_\mathrm{ref}(y|x)$, adjusted by the relative likelihood ratio $\left(\frac{p_\mathcal{D}(y|x)}{q_\mathcal{D}(y|x)}\right)^{1/\beta}$.

In most recommendation settings, negative samples are uniformly distributed \cite{DMPO, chen2024softmax}, i.e., $q_\mathcal{D}(y|x) = \mathcal{U} = \frac{1}{|\mathcal{I}|}$, where $\mathcal{I}$ is the set of all candidate items. Additionally, in typical DPO-based preference alignment scenarios, $\beta$ is constrained to $0 < \beta < 1$. Under these conditions, \emph{the DPO loss inherently biases the model toward popular items with higher $p_\mathcal{D}(y|x)$, exacerbating popularity bias.}
In the extreme case where $\beta \to 0$, the optimal policy collapses to recommending only the most popular items, effectively disregarding less frequent but potentially valuable recommendations.



\medskip \noindent
\textbf{Remark:} 
This result is a byproduct of DPO's loss function. Unlike the forward KL-divergence $\mathbb{D}_{KL}(p_\mathcal{D}(y|x), \pi_\theta(y|x))$ used in the SFT loss in \myeq{forward_KL}, DPO optimizes the reverse KL-divergence $\mathbb{D}_{KL}(\pi_\theta(y|x), \pi_\mathrm{ref}(y|x))$. Forward KL-divergence is known for its mass-covering property, which encourages learning an average behavior and is less sensitive to subtle differences in the preference distribution (as illustrated in \myfig{method}(a)) \cite{sun2024inverse}. In contrast, the reverse KL-divergence used in DPO promotes mode-seeking behavior \cite{wang2023beyond, omura2024entropy}, guiding the model to focus on the ``peaks'' of the distribution (\myfig{method}(b)).

The issue of DPO has also been highlighted by other researchers \cite{pal2024smaug}. Specifically, \citet{feng2024towards} derive that DPO suppresses negative samples more aggressively than it elevates positive samples during optimization. Moreover, \citet{IPO} demonstrate that the empirical optimal policy often drives $\pi_\theta(y_l|x) \rightarrow 0$ for all $\beta$, stemming from an underfitting of the potential reward.

In the context of recommendation, this behavior can be detrimental, as it exacerbates the filter bubble issue and undermines user interests by limiting exposure to diverse items \cite{gao2023alleviating}.

\section{Method}
\label{sec:method}
We present how to address the popularity bias in DPO by utilizing the self-play philosophy. Then we detail the proposed SPRec architecture.

\subsection{Solution: Suppress Biases through Self-Play}
\label{sec:solution}

Since the DPO loss inherently causes the policy $\pi_\theta$ to learn sharp ``peaks'', leading to bias, an intuitive solution is to directly suppress these learned peaks. To address this, we utilize a self-play framework, dubbed SPDPO, which iteratively alternates between policy learning and bias suppression. Specifically, in the $(t+1)$-th iteration, negative samples are drawn from the model's predictive distribution $\pi_{\theta_t}(\cdot|x)$ at iteration $t$, resulting in the following learning paradigm:

\begin{equation}
\label{eq:SPDPO}
\begin{aligned} 
    \pi_{\theta_{t+1}} \leftarrow \arg\max_{\pi_\theta} \mathbb{E}_{(x,y_w)\sim D, y_l\sim \pi_{\theta_t}(\cdot|x)}l(\pi_\theta;\pi_{\theta_t};x,y_w,y_l).
\end{aligned}
\end{equation}
By comparing \myeq{SPDPO} with \myeq{dpo_new}, we obtain that the objective function $\mathcal{L}_{SPDPO}$ in the $(t+1)$-th iteration can be viewed as $\mathcal{L}_{DPO}$ weighted by $\frac{\pi_{\theta_t}(y_l|x)}{q_\mathcal{D}(y_l|x)}$, which can be expressed as follows:
\begin{equation}
\label{eq:SPDPO2}
\begin{aligned} 
    \mathcal{L}_{SPDPO} = -\mathbb{E}_{(x,y_w)\sim \mathcal{D}, y_l\sim q_\mathcal{D}(\cdot|x)} \frac{\pi_{\theta_t}(y_l|x)}{q_\mathcal{D}(y_l|x)}l(\pi_\theta;\pi_{\theta_t};x,y_w,y_l).
\end{aligned}
\end{equation}
Again, if DPO uses negative samples from a discrete uniform distribution $q_\mathcal{D}(y|x)=\mathcal{U}=\frac{1}{|\mathcal{I}|}$, then the objective function $\mathcal{L}_{SPDPO}$ can be viewed as $\mathcal{L}_{DPO}$ weighted by $\pi_{\theta_t}(y_l|x)$.
This highlights that the objective adaptively pays more attention to biased items by increasing their learning rates if they have higher probabilities in the model's output distribution. 

\medskip \noindent
\textbf{Remark:} 
Unlike traditional recommendation methods that predefine negative samples or allocate weights in advance, our approach dynamically selects negative samples during the learning process. This provides a \emph{significant advantage, enabling the model to adaptively adjust its learning paradigm for effective bias suppression}. As a result, this approach mitigates the filter bubble issue and enhances the diversity of recommendations.

\subsection{Architecture of SPRec}
Utilizing the loss function in \myeq{SPDPO}, we propose a self-play recommendation tuning framework, SPRec, which generally includes multiple iterations of both an SFT step and a DPO step. The workflow is illustrated in \myfig{method}(c), in which three key steps are conducted sequentially in each iteration:
\begin{enumerate}[leftmargin=*]
    \item \textbf{Dataset Construction}: For each positive sample $\{(x^i, y_w^i)\}$ in the offline dataset, sample a negative sample $y_l^i$ by running the current model $\pi_{\theta_t}$ and using its predicted recommendation as $y_l^i$. Thus we obtain pairwise preference data for each sample as $\{(x^i, y_w^i, y_l^i)\}$.
    \item \textbf{SFT Step}: Use only the positive sample $\{(x^i, y_w^i)\}$ to refine the model $\pi_{\theta_t}$ through SFT techniques such as instruction learning. 
    \item \textbf{DPO Step}: Align the model $\pi_{\theta_t}$ by perform DPO step using pairwise dataset $\{(x^i, y_w^i, y_l^i)\}$, and obtain $\pi_{\theta_{t+1}}$.
\end{enumerate}

This process repeats for $T$ iterations per epoch. The self-play mechanism is adaptable to any LLM-based recommender system. 
To ensure comparability with existing DPO-based recommenders \cite{chen2024softmax, DMPO, liao2024rosepo}, we can extend from a single to multiple negative samples, with results analyzed in experiments. 



\medskip \noindent
\textbf{Remark:} 
Although the loss function in \myeq{SPDPO} is inherently capable of aligning the model with positive samples, our experiments reveal that incorporating an SFT step in each self-play iteration can further enhance performance.

In fact, combining SFT and DPO has been shown to be an effective practice in recent research and open-sourced LLM models. For example, each iteration of post-training for Llama 3 includes an SFT stage followed by a DPO stage \cite{llama3}. Similarly, \citet{pang2024iterative} demonstrate the effectiveness of an iterative preference optimization algorithm using a modified DPO loss with an additional negative log-likelihood (NLL) term, which mirrors the SFT loss defined in \myeq{sft}.

\section{Experiments}
In this section, we conduct experiments to address the following research questions:

\begin{itemize}[leftmargin=10pt]
    \item \textbf{RQ1}: How does the SPRec training framework compare to baseline methods in terms of accuracy, diversity, and fairness?
    \item \textbf{RQ2}: What are the contributions of different components within the SPRec framework?
    \item \textbf{RQ3}: How do the random sampling ratio and the number of negative samples impact the performance?
\end{itemize}

\begin{table}[!t]
\centering
\caption{
Overall performance comparison of \emphcolorgreenn{SPRec (green)}, \emphcolorbrownn{SFT-based (brown)}, and \emphcolorbluee{DPO-based (blue)} methods. Best results are bold, sub-optimal ones underlined. $\uparrow$ indicates higher is better, while $\downarrow$ indicates lower is better.}
\label{tab:overall}
\tabcolsep=1pt
\small
\renewcommand\arraystretch{0.9}
\begin{tabular}{cccccccc}
\toprule
\textbf{Dataset} &
  \textbf{Model} &
  \textbf{DivRatio$\uparrow$} &
  \textbf{ORRatio$\downarrow$} &
  \textbf{MGU$\downarrow$} &
  \textbf{HR$\uparrow$} &
  \textbf{NDCG$\uparrow$} \\ \midrule
 &
  SASRec &
  0.0031 &
  1.0000 &
  0.1209 &
  0.0225 &
  0.0136 \\
\cdashlinelr{2-8}
 &
  \cellcolor{brown!15}BIGRec &
  \cellcolor{brown!15}{\ul 0.1939} &
  \cellcolor{brown!15}0.2561 &
  \cellcolor{brown!15}0.0620 &
  \cellcolor{brown!15}{\ul 0.0347} &
  \cellcolor{brown!15}{\ul 0.0281} \\
 &
  \cellcolor{brown!15}RW &
  \cellcolor{brown!15}0.1918 &
  \cellcolor{brown!15}0.2551 &
  \cellcolor{brown!15}0.0577 &
  \cellcolor{brown!15}0.0327 &
  \cellcolor{brown!15}0.0276 \\
 &
  \cellcolor{brown!15}$\text{D}^3$ &
  \cellcolor{brown!15}0.1246 &
  \cellcolor{brown!15}0.3238 &
  \cellcolor{brown!15}0.0664 &
  \cellcolor{brown!15}0.0266 &
  \cellcolor{brown!15}0.0197 \\
\cdashlinelr{2-8}
 &
  \cellcolor{gblue!10}DMPO &
  \cellcolor{gblue!10}0.1827 &
  \cellcolor{gblue!10}0.2561 &
  \cellcolor{gblue!10}0.0529 &
  \cellcolor{gblue!10}0.0310 &
  \cellcolor{gblue!10}0.0264 \\
  &
  \cellcolor{gblue!10}SDPO &
  \cellcolor{gblue!10}0.1816 &
  \cellcolor{gblue!10}0.2449 &
  \cellcolor{gblue!10}{\ul 0.0462} &
  \cellcolor{gblue!10}0.0310 &
  \cellcolor{gblue!10}0.0258 \\
 &
  \cellcolor{gblue!10}RosePO &
  \cellcolor{gblue!10}0.1857 &
  \cellcolor{gblue!10}{\ul 0.2378} &
  \cellcolor{gblue!10}0.0538 &
  \cellcolor{gblue!10}0.0290 &
  \cellcolor{gblue!10}0.0244 \\
\cdashlinelr{2-8}
\multicolumn{1}{c}{\multirow{-8}{*}{\textbf{MovieLens}}} &
  \cellcolor{green!8}SPRec &
  \cellcolor{green!8}\textbf{0.2806} &
  \cellcolor{green!8}\textbf{0.1510} &
  \cellcolor{green!8}\textbf{0.0432} &
  \cellcolor{green!8}\textbf{0.0388} &
  \cellcolor{green!8}\textbf{0.0319} \\ \midrule
 &
  SASRec &
  0.0030 &
  1.0000 &
  0.0458 &
  0.0202 &
  0.0139 \\
\cdashlinelr{2-8}
 &
  \cellcolor{brown!15}BIGRec &
  \cellcolor{brown!15}0.1420 &
  \cellcolor{brown!15}0.3170 &
  \cellcolor{brown!15}0.0175 &
  \cellcolor{brown!15}0.0310 &
  \cellcolor{brown!15}0.0236 \\
 &
  \cellcolor{brown!15}RW &
  \cellcolor{brown!15}0.1050 &
  \cellcolor{brown!15}0.3730 &
  \cellcolor{brown!15}0.0238 &
  \cellcolor{brown!15}0.0380 &
  \cellcolor{brown!15}0.0281 \\
   &
  \cellcolor{brown!15}$\text{D}^3$ &
  \cellcolor{brown!15}0.1199 &
  \cellcolor{brown!15}{\ul 0.2581} &
  \cellcolor{brown!15}0.0260 &
  \cellcolor{brown!15}{\ul 0.0413} &
  \cellcolor{brown!15}\textbf{0.0324} \\
\cdashlinelr{2-8}
 &
  \cellcolor{gblue!10}DMPO &
  \cellcolor{gblue!10}0.1560 &
  \cellcolor{gblue!10}0.3310 &
  \cellcolor{gblue!10}0.0164 &
  \cellcolor{gblue!10}0.0410 &
  \cellcolor{gblue!10}0.0314 \\
  &
  \cellcolor{gblue!10}SDPO &
  \cellcolor{gblue!10}0.1580 &
  \cellcolor{gblue!10}0.3270 &
  \cellcolor{gblue!10}{\ul 0.0161} &
  \cellcolor{gblue!10}\textbf{0.0420} &
  \cellcolor{gblue!10}{\ul 0.0315} \\
 &
  \cellcolor{gblue!10}RosePO &
  \cellcolor{gblue!10}{\ul 0.1860} &
  \cellcolor{gblue!10}0.3230 &
  \cellcolor{gblue!10}0.0181 &
  \cellcolor{gblue!10}0.0300 &
  \cellcolor{gblue!10}0.0215 \\
\cdashlinelr{2-8}
\multicolumn{1}{c}{\multirow{-8}{*}{\textbf{Goodreads}}} & 
  \cellcolor{green!8}SPRec &
  \cellcolor{green!8}\textbf{0.2090} &
  \cellcolor{green!8}\textbf{0.2170} &
  \cellcolor{green!8}\textbf{0.0099} &
  \cellcolor{green!8}0.0330 &
  \cellcolor{green!8}0.0250 \\ \midrule
 &
  SASRec &
  0.0021 &
  1.0000 &
  0.1205 &
  0.0226 &
  0.0170 \\
\cdashlinelr{2-8}
 &
  \cellcolor{brown!15}BIGRec &
  \cellcolor{brown!15}0.3198 &
  \cellcolor{brown!15}0.2597 &
  \cellcolor{brown!15}0.0461 &
  \cellcolor{brown!15}{\ul 0.0132} &
  \cellcolor{brown!15}{\ul 0.0130} \\
 &
  \cellcolor{brown!15}RW &
  \cellcolor{brown!15}0.2821 &
  \cellcolor{brown!15}0.2648 &
  \cellcolor{brown!15}0.0432 &
  \cellcolor{brown!15}0.0112 &
  \cellcolor{brown!15}0.0103 \\
 &
  \cellcolor{brown!15}$\text{D}^3$ &
  \cellcolor{brown!15}0.3135 &
  \cellcolor{brown!15}0.2672 &
  \cellcolor{brown!15}0.0320 &
  \cellcolor{brown!15}0.0103 &
  \cellcolor{brown!15}0.0088 \\
\cdashlinelr{2-8}
 &
  \cellcolor{gblue!10}DMPO &
  \cellcolor{gblue!10}0.3116 &
  \cellcolor{gblue!10}{\ul 0.1740} &
  \cellcolor{gblue!10}\textbf{0.0195} &
  \cellcolor{gblue!10}0.0090 &
  \cellcolor{gblue!10}0.0090 \\
 &
  \cellcolor{gblue!10}SDPO &
  \cellcolor{gblue!10}0.3218 &
  \cellcolor{gblue!10}0.2118 &
  \cellcolor{gblue!10}0.0380 &
  \cellcolor{gblue!10}0.0110 &
  \cellcolor{gblue!10}0.0106 \\
 &
  \cellcolor{gblue!10}RosePO &
  \cellcolor{gblue!10}{\ul 0.3625} &
  \cellcolor{gblue!10}0.1853 &
  \cellcolor{gblue!10}0.0426 &
  \cellcolor{gblue!10}0.0090 &
  \cellcolor{gblue!10}0.0090 \\
\cdashlinelr{2-8}
\multicolumn{1}{c}{\multirow{-8}{*}{\textbf{CDs\_and\_Vinyl}}} &
  \cellcolor{green!8}SPRec &
  \cellcolor{green!8}\textbf{0.3859} &
  \cellcolor{green!8}\textbf{0.1670} &
  \cellcolor{green!8}{\ul 0.0242} &
  \cellcolor{green!8}\textbf{0.0143} &
  \cellcolor{green!8}\textbf{0.0140} \\ \midrule
 &
  SASRec &
  0.0010 &
  1.0000 &
  0.1094 &
  0.0660 &
  0.0379 \\
\cdashlinelr{2-8}
 &
  \cellcolor{brown!15}BIGRec &
  \cellcolor{brown!15}0.1940 &
  \cellcolor{brown!15}0.3910 &
  \cellcolor{brown!15}0.0650 &
  \cellcolor{brown!15}0.0780 &
  \cellcolor{brown!15}0.0766 \\
 &
  \cellcolor{brown!15}RW &
  \cellcolor{brown!15}{\ul 0.2890} &
  \cellcolor{brown!15}{\ul 0.2710} &
  \cellcolor{brown!15}\textbf{0.0313} &
  \cellcolor{brown!15}0.0760 &
  \cellcolor{brown!15}0.0735 \\
 &
  \cellcolor{brown!15}$\text{D}^3$ &
  \cellcolor{brown!15}0.1580 &
  \cellcolor{brown!15}0.4560 &
  \cellcolor{brown!15}0.0511 &
  \cellcolor{brown!15}0.0730 &
  \cellcolor{brown!15}0.0718 \\
\cdashlinelr{2-8}
 &
  \cellcolor{gblue!10}DMPO &
  \cellcolor{gblue!10}0.2270 &
  \cellcolor{gblue!10}0.2990 &
  \cellcolor{gblue!10}0.0443 &
  \cellcolor{gblue!10}{\ul 0.0850} &
  \cellcolor{gblue!10}{\ul 0.0834} \\  
  &
  \cellcolor{gblue!10}SDPO &
  \cellcolor{gblue!10}0.2080 &
  \cellcolor{gblue!10}0.3510 &
  \cellcolor{gblue!10}0.0475 &
  \cellcolor{gblue!10}0.0820 &
  \cellcolor{gblue!10}0.0810 \\
 &
  \cellcolor{gblue!10}RosePO &
  \cellcolor{gblue!10}0.2310 &
  \cellcolor{gblue!10}0.3160 &
  \cellcolor{gblue!10}0.0499 &
  \cellcolor{gblue!10}0.0820 &
  \cellcolor{gblue!10}0.0805 \\
\cdashlinelr{2-8}
\multicolumn{1}{c}{\multirow{-8}{*}{\textbf{Steam}}}  &
  \cellcolor{green!8}SPRec &
  \cellcolor{green!8}\textbf{0.2930} &
  \cellcolor{green!8}\textbf{0.2560} &
  \cellcolor{green!8}{\ul 0.0367} &
  \cellcolor{green!8}{\ul \textbf{0.0910}} &
  \cellcolor{green!8}\textbf{0.0893} \\ \bottomrule
\end{tabular}%
\end{table}

\subsection{Experimental Setup}

\subsubsection{Datasets}
\label{sec:datasets}
We conducted extensive experiments on four real-world datasets: \emph{MovieLens}\footnote{\url{https://grouplens.org/datasets/movielens/}}, \emph{Steam}\footnote{\url{https://cseweb.ucsd.edu/\~jmcauley/datasets.html\#amazon\_reviews}}, \emph{Goodreads}\footnote{\url{https://mengtingwan.github.io/data/goodreads}}, and the \emph{CDs and Vinyl} category of the Amazon Review Dataset\footnote{\url{http://jmcauley.ucsd.edu/data/amazon/index\_2014.html}}. 
Additional details about the datasets are provided in the \myappendix{datasets}.
Following the data processing approach in \cite{chen2024softmax, bao2024decoding}, interaction sequences with fewer than 10 entries were excluded. The datasets were then split chronologically into training, validation, and test sets in an 8:1:1 ratio, ensuring mutual exclusivity and preventing data leakage. 
To ensure comparability across different LLM-based methods, we further sampled 4,096 interactions from each dataset's training set as the training samples for all methods, 512 interactions from the validation set, and 1,000 interactions from the test set.


To process category information, we extracted category metadata from each dataset and identified the most $10$ popular categories within the training sets. To ensure category independence, we removed categories with clear hierarchical relationships, such as ``FPS'' and ``Shooting'' in the Steam dataset, and ``Rock'' and ``Classical Rock'' in the CDs and Vinyl dataset.

\begin{figure*}[h]
  \centering
  \includegraphics[width=\linewidth]{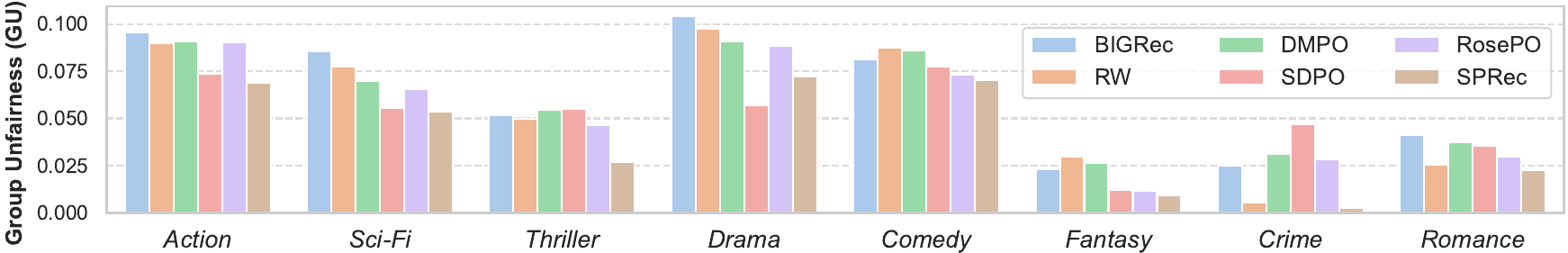}
   \caption{Comparison of models across genres on Group Unfairness (GU) in top-$1$ recommendation.}
 \Description{Fig: The Group Unfairness (GU) of different groups divided by genres in top-1 recommendation.}
  \label{fig:GU}
\end{figure*}


    
    
    

\subsubsection{Evaluation Setting}
To leverage the strengths of LLMs in generative recommendation tasks, we prompt the LLM to generate a predicted item based on the input history sequence. Then, following the procedures in BIGRec \cite{bao2023bi}, we calculate scores and rankings for the entire item space and ground our predicted item to an exact item in the dataset.

\subsubsection{Metrics.}  
We evaluate the model on 1,000 randomly sampled test cases per iteration using four key metrics. Accuracy is measured by NDCG@5 and HR@5, averaged across results. Diversity is assessed via DivRatio, representing the proportion of unique recommendations. Over-recommendation is quantified by ORRatio, indicating the proportion of results dominated by the three most frequently recommended items. Fairness is evaluated using MGU \cite{jiang2024item}, capturing category-level discrepancies between recommendations and user history.

\subsubsection{Baseline}
For traditional recommendation models, we select SASRec \cite{kang2018self}, a widely used baseline employing a sequential method with a self-attention mechanism. 
For LLM-based models, we consider several baselines. 
(1) For SFT-based methods, \textbf{BIGRec} \cite{bao2023bi} serves as an instruction-tuning LLM framework for sequential recommendations and forms the foundation for SPRec. \textbf{Re-weighting (RW)} \cite{jiang2024item} improves fairness in BIGRec by balancing recommendations across categories through dataset-based training weights. \textbf{Debiasing-Diversifying Decoding ($\textbf{D}^3$)} \cite{bao2024decoding} enhances diversity in BIGRec using a decoding strategy guided by SASRec. 
(2) For DPO-based models, \textbf{DMPO} \cite{DMPO} introduces DPO into LRSs by sampling multiple negative items as rejected responses, while \textbf{Softmax-DPO (SDPO)} \cite{chen2024softmax} follows a similar approach but incorporates a softmax loss over multiple negative samples. Finally, \textbf{RosePO} \cite{liao2024rosepo} is a preference optimization framework that combines negative sampling strategies and personalized uncertainty to achieve fairness, unbiasedness, and robustness.
The implementation details are listed in \myappendix{details}.

\subsection{Overall Performance Comparison (RQ1)}  
\label{sec:RQ1}  


The experimental results are presented in \mytab{overall}, leading to the following observations. The non-LLM baseline, SASRec, performs poorly with the given training size, which is expected as SASRec requires large datasets to achieve effective fitting. In this study, we primarily focus on LLM-based methods, and SASRec's results are included only for reference and as the assistant model for $\text{D}^3$ during the decoding stage.

\subsubsection{Limitations of SFT-based Methods}  
Fine-tuning LLMs with instruction-based methods results in recommendations heavily favoring popular items, leading to a lack of diversity. For example, in the Goodreads dataset, the DivRatio of BIGRec is only 0.142, meaning the model provides just 14 distinct recommendations per 100 tasks. Similarly, in the Steam dataset, BIGRec's ORRatio reaches 0.391, with over 39\% of recommendations concentrated on the 3 most popular items. These findings highlight that relying solely on SFT introduces severe biases, significantly overexposing certain popular items.  

\subsubsection{Limitations of DPO-based Methods}  
For DPO methods using random sampling, such as SDPO and DMPO, while multiple negative samples improve recommendation accuracy, they perform poorly on diversity and fairness metrics. On the Goodreads and MovieLens datasets, SDPO and DMPO have minimal impact on DivRatio and ORRatio and may even degrade model performance. On the CDs and Steam datasets, although ORRatio decreases, diversity metrics remain largely unchanged, suggesting that the model favors moderately popular items but fails to effectively recommend new ones. In contrast, RosePO performs well on the CD dataset due to its negative sampling strategy based on semantic information. However, this approach heavily relies on the semantic characteristics of the dataset's structure, resulting in relatively poor performance on other datasets and limiting its generalizability for debiasing.

In summary, existing DPO-based methods fail to address fairness issues in LRS.

\subsubsection{Superiority of SPRec}  
As shown in \mytab{overall}, SPRec significantly improves both DivRatio and ORRatio metrics across all datasets compared to BIGRec, demonstrating its effectiveness in mitigating the over-recommendation of popular items and enhancing diversity. Additionally, SPRec outperforms BIGRec on most fairness metrics, reducing the discrepancies between the model's recommendations and users' historical sequences, thereby providing fairer recommendations.  

SPRec also surpasses all baseline models on DivRatio and ORRatio, showcasing its superior ability to balance recommendation distributions. For fairness, SPRec achieved the highest MGU scores on the MovieLens and Goodreads datasets, and the second-highest on the Steam and CD datasets. Moreover, as shown in \myfig{GU}, SPRec alleviates category-level unfairness on the MovieLens dataset, achieving the best results in 7 out of 8 categories, further underscoring its effectiveness in improving fairness.  

While RosePO performs well on the CDs and Vinyl dataset, leveraging semantic-based negative sampling to address fairness in music recommendations, and Re-weighting shows strong performance on the Steam dataset by employing category-based re-weighting for gaming recommendations, these methods are tailored to specific datasets and lack generalizability. In contrast, SPRec’s self-play framework provides a universal solution, overcoming dataset-specific challenges and delivering fairer recommendations across diverse scenarios.


\begin{table}[!t]
\centering
\caption{Ablation results. ``with RN'' for random negative samples, ``w/o'' for without specific components.}
\label{tab:ablation}
\tabcolsep=1pt
\small
\renewcommand\arraystretch{0.9}
\begin{tabular}{cccccccccc}
\toprule
\textbf{Dataset}                                      & \textbf{Model}               & \textbf{DivRatio$\uparrow$}            & \textbf{ORRatio$\downarrow$}               & \textbf{MGU$\downarrow$}                                              & \textbf{HR$\uparrow$}                         & \textbf{NDCG$\uparrow$}                                  \\ \midrule
\multicolumn{1}{c}{}                                  & w/o SFT                  & \textbf{0.3020}                          & \textbf{0.0837}                              & \textbf{0.0198}                           & 0.0184                                  & 0.0149                                  \\
\multicolumn{1}{c}{}                                  & w/o DPO                  & 0.1959                                   & 0.2714                                       & 0.0637                                    & {\ul 0.0316}                            & {\ul 0.0260}                            \\
\multicolumn{1}{c}{}                                  & with RN                     & 0.2194                                   & 0.2224                                       & 0.0544                                    & 0.0286                                  & 0.0230                                  \\
\multicolumn{1}{c}{\multirow{-4}{*}{\textbf{MovieLens}}}       & \cellcolor{green!8}SPRec & \cellcolor{green!8}{\ul 0.2806}     & \cellcolor{green!8}{\ul 0.1510}              & \cellcolor{green!8}{\ul 0.0432}      & \cellcolor{green!8}\textbf{0.0388} & \cellcolor{green!8}\textbf{0.0319} \\ \midrule
\multicolumn{1}{c}{}                                  & w/o SFT                  & {\ul 0.2010}                             & {\ul 0.2390}                                 & \textbf{0.0044}                           & 0.0270                                  & 0.0206                                  \\
\multicolumn{1}{c}{}                                  & w/o DPO                  & 0.1350                                   & 0.2970                                       & 0.0142                                    & {\ul 0.0350}                            & {\ul 0.0274}                            \\
\multicolumn{1}{c}{}                                  & with RN                     & 0.1380                                   & 0.3380                                       & 0.0188                                    & \textbf{0.0420}                         & \textbf{0.0310}                         \\
\multicolumn{1}{c}{\multirow{-4}{*}{\textbf{Goodreads}}}       & \cellcolor{green!8}SPRec & \cellcolor{green!8}\textbf{0.2090}  & \cellcolor{green!8}\textbf{0.2170}           & \cellcolor{green!8}{\ul 0.0099}      & \cellcolor{green!8}0.0330          & \cellcolor{green!8}0.0250          \\ \midrule
\multicolumn{1}{c}{}                                  & w/o SFT                  & 0.3381                                   & 0.2373                                       & \textbf{0.0216}                           & 0.0132                                  & 0.0126                                  \\
\multicolumn{1}{c}{}                                  & w/o DPO                  & 0.3136                                   & {\ul 0.2363}                                 & 0.0333                                    & 0.0143                                  & 0.0136                                  \\
\multicolumn{1}{c}{}                                  & with RN                     & {\ul 0.3625}                             & 0.2536                                       & 0.0359                                    & \textbf{0.0163}                         & \textbf{0.0150}                         \\
\multicolumn{1}{c}{\multirow{-4}{*}{\textbf{CDs\_and\_Vinyl}}} & \cellcolor{green!8}SPRec & \cellcolor{green!8}\textbf{0.3859}  & \cellcolor{green!8}\textbf{0.1670}           & \cellcolor{green!8}{\ul 0.0242}      & \cellcolor{green!8}{\ul 0.0143}    & \cellcolor{green!8}{\ul 0.0140}    \\ \midrule
\multicolumn{1}{c}{}                                  & w/o SFT                  & {\ul 0.2900}                             & \textbf{0.2260}                              & \textbf{0.0173}                           & {\ul 0.0880}                            & {\ul 0.0868}                            \\
\multicolumn{1}{c}{}                                  & w/o DPO                  & 0.2220                                   & 0.3910                                       & 0.0620                                    & 0.0790                                  & 0.0776                                  \\
\multicolumn{1}{c}{}                                  & with RN                     & 0.2860                                   & {\ul 0.2530}                                 & {\ul 0.0351}                              & 0.0860                                  & 0.0837                                  \\
\multicolumn{1}{c}{\multirow{-4}{*}{\textbf{Steam}}}           & \cellcolor{green!8}SPRec & \cellcolor{green!8}\textbf{0.2930}  & \cellcolor{green!8}0.2560                    & \cellcolor{green!8}0.0367            & \cellcolor{green!8}\textbf{0.0910} & \cellcolor{green!8}\textbf{0.0893} \\ \bottomrule
\end{tabular}%
\end{table}

\begin{figure}[!t]
  \centering
   \includegraphics[width=\linewidth]{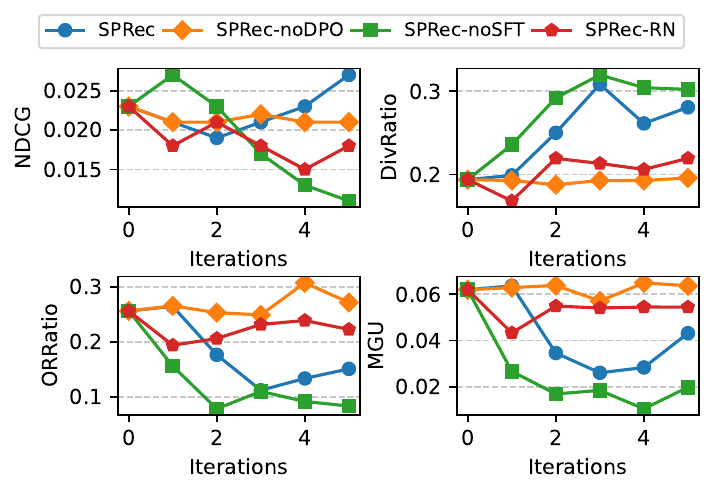}
  \caption{Performance on the MovieLens dataset across different ablation experiments.}
  \label{fig:ablation}
\end{figure}

\subsection{Ablation Study (RQ2)}
We conducted a series of ablation experiments to explore the impact of each component of the SPRec training framework.

\subsubsection{SPRec without SFT}
As shown in \mytab{ablation}, SPRec w/o SFT achieves the lowest recommendation accuracy across all datasets except Steam. This indicates that, during the self-play process, the model's excessive focus on fairness compromises its accuracy. In the MovieLens dataset (\myfig{ablation}), the absence of the SFT stage leads to a steady decline in recommendation accuracy (NDCG) throughout training. These findings highlight the critical role of SFT in maintaining SPRec's recommendation performance.

\subsubsection{SPRec without DPO}
Removing DPO reduces SPRec to further SFT training, ensuring that performance gains are not due to incorporating additional data. As shown in \mytab{ablation}, additional SFT fails to improve diversity or fairness metrics, and the recommendations remain biased. Furthermore, \myfig{ablation} reveals minimal fluctuations during training, indicating that the prior SFT training has already converged. This experiment underscores the limitations of SFT-based methods in addressing recommendation fairness and diversity.

\subsubsection{Randomly sampling negative items}
As observed in \mytab{ablation}, when the negative sampling strategy is replaced with random sampling, SPRec-RN fails to achieve further improvements in DivRatio and ORRatio metrics on the MovieLens and Goodreads datasets. Additionally, SPRec-RN's fairness metrics perform worse compared to SPRec. Although SPRec-RN shows a significant improvement in DivRatio on the CDs and Vinyl dataset, its ORRatio still performs poorly. This suggests that random sampling of negative samples during training is ineffective at suppressing popular items, and the recommendation results continue to exhibit a significant long-tail effect. This ablation experiment demonstrates that our Self-play negative sampling strategy effectively balances the distribution of the model's output, leading to debiasing in recommendations.
Replacing the negative sampling strategy with random sampling (SPRec-RN) fails to improve DivRatio and ORRatio metrics on the MovieLens and Goodreads datasets (\mytab{ablation}). Additionally, SPRec-RN exhibits worse fairness metrics compared to SPRec. While it achieves a significant boost in DivRatio on the CDs dataset, its ORRatio remains poor. These results suggest that random negative sampling is ineffective in suppressing popular items, leaving a pronounced long-tail effect in the recommendations. This experiment demonstrates that our self-play negative sampling strategy effectively balances the model's output distribution, resulting in debiased recommendations.

 \begin{figure}[!t]
  \centering
  \setlength{\abovecaptionskip}{0.1cm}
 \setlength{\belowcaptionskip}{0cm}
  \includegraphics[width=\linewidth]{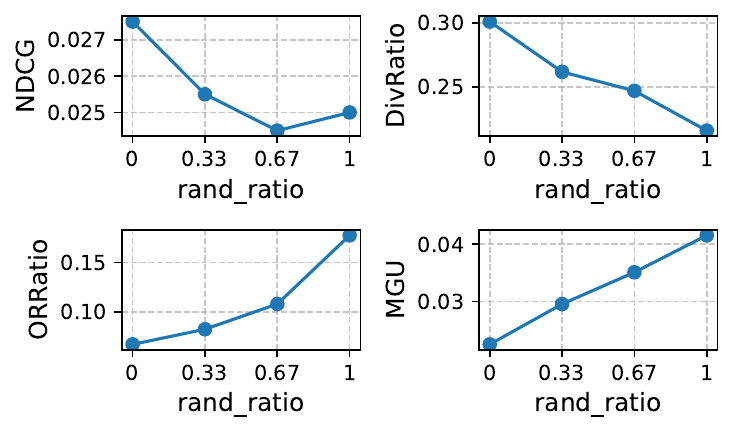}
  \caption{Effect of random sampling ratio.}
  \label{fig:RQ3_rand_ratio}
\end{figure}

 \begin{figure}[!t]
  \centering
  \setlength{\abovecaptionskip}{0.1cm}
  \setlength{\belowcaptionskip}{0cm}
  \includegraphics[width=\linewidth]{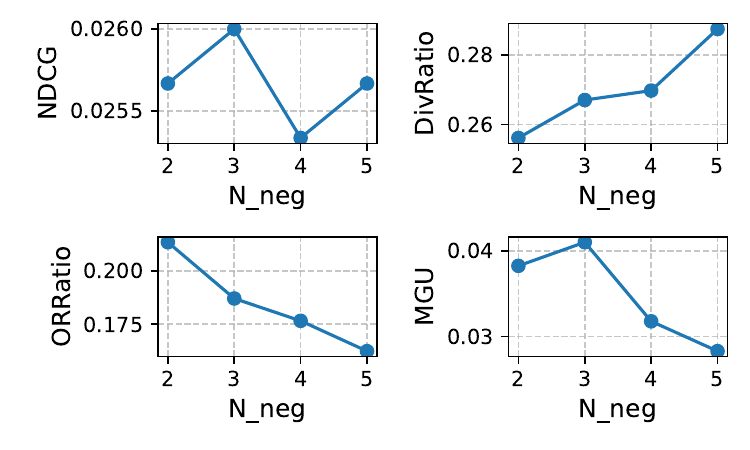}
  \vspace{-3mm}
  \caption{Effect of negative sample size.}
  \label{fig:RQ3_N_neg}
  \vspace{-2mm}
\end{figure}

\subsection{Impact of Negative Samples (RQ3)}

We investigate the role of negative samples in SPRec by introducing a proportion of random negative samples to contaminate SPRec's original self-play samples. Additionally, we examine the impact of increasing the number of negative samples in SPRec’s loss function (\myeq{SPDPO2}). To achieve this, we adopt the SDPO loss function to expand to multiple negative samples. For efficiency, we limit the training sample size to 1,024, keeping other experimental settings consistent with \mysec{RQ1}. To ensure result stability, the model's performance is averaged over the last three training iterations.  We report the results on the Movielens dataset. 

\subsubsection{Effect of Random Sampling Ratio}
In each training iteration, we randomly replace a proportion of negative samples with randomly selected items while leaving the remaining negative samples unchanged as the model's recommendation outputs. As shown in \myfig{RQ3_rand_ratio}, increasing the proportion of random negative samples leads to a steady decline in diversity and accuracy. Fairness also deteriorates, with recommendations becoming more skewed toward a small subset of popular items. These results highlight the superiority of our proposed self-play negative sampling strategy over random sampling.

\subsubsection{Effect of Negative Sample Size}
To generate $N$ negative samples, we use beam search decoding to sample $2N$ items from the model's output. After deduplication, the top $N$ items with the highest predicted probabilities are selected as negative samples. As shown in \myfig{RQ3_N_neg}, increasing the number of negative samples results in stable recommendation accuracy but significantly improves diversity and fairness, reducing the focus on popular items. This demonstrates the versatility of the self-play negative sampling strategy, which can be effectively combined with multi-negative sampling approaches to further debias LRS.



\section{Conclusion \& Discussion}

Our work establishes a critical bridge between preference alignment techniques and fairness-aware recommendation in the era of LLMs. Through both theoretical analysis and empirical validation, we demonstrate that conventional DPO-based tuning fundamentally conflicts with the principles of equitable recommendation, creating self-reinforcing popularity biases that traditional debiasing approaches fail to address. The proposed SPRec framework represents a paradigm shift in recommendation alignment - rather than treating bias mitigation as a post-hoc correction, we redesign the core learning mechanism to enable autonomous bias suppression through self-regulated competition between model generations. This approach not only achieves state-of-the-art performance across accuracy and fairness metrics but more importantly, provides a blueprint for developing self-correcting AI systems that maintain alignment with both user preferences and ethical constraints.


Despite its effectiveness, our work primarily addresses bias in DPO-based tuning, while overlooking the popularity bias already present in SFT due to its cross-entropy loss. Future research should focus on mitigating bias at the SFT stage to ensure fairness from the start of fine-tuning.
Additionally, optimizing preferences in recommendation is a long-term challenge, requiring alignment across sequential recommendations rather than individual predictions. However, LLMs generate outputs token by token, making it difficult to optimize preferences from token-level to item-level and ultimately list-level recommendations. Tackling this issue will require new datasets, benchmarks, and models capable of long-term alignment. A promising direction is reinforcement learning with process-level rewards, shifting optimization from short-term token likelihood to long-horizon user engagement.

\section*{Acknowledgements}
This work is supported by the National Key Research and Development Program of China (2021ZD0111802), the National Natural Science Foundation of China (62402470, 62272437, U24B20180, 62121002), the Fundamental Research Funds for the Central Universities of China (WK2100000053, PA2024GDSK0107), Anhui Provincial Natural Science Foundation (2408085QF189), and the Postdoctoral Fellowship Program of CPSF (GZC20241643). This research is supported by the advanced computing resources provided by the Supercomputing Center of the USTC.

\bibliographystyle{ACM-Reference-Format}
\balance
\bibliography{SPRec}

\appendix

\section{Mathematical Derivations}
\label{appendix:DPO}
\begin{proof}[Proof of \mytheorem{dpo_theorem}]
The DPO loss is derived from the objective of Reinforcement Learning with Human Feedback (RLHF):
\begin{equation}\label{eqn:rlhf_def}
\max_\theta \mathbb E_{x\sim \mathcal D, y\sim\pi_\theta(\cdot|x)}[r(x,y)] - \beta \mathrm{D_{KL}}(\pi_\theta\|\pi_\mathrm{ref}),
\end{equation}
where the reward model is defined via the BT model:
$$
\max_r \mathbb E_{(x,y_w)\sim\mathcal D, y_l\sim Y_u}\log\sigma(r(x,y_w) - r(x,y_l)).
$$

In the original paper of DPO \cite{rafailov2024direct}, the authors proved that the optimal policy $\pi_\theta^*$ for DPO loss in Eq.~\myeq{dpo_new} and the solution to the optimization problem in Eq.~\eqref{eqn:rlhf_def} are the same. Thus, we can analyze the solution to Eq.~\eqref{eqn:rlhf_def}, equivalent to examining the DPO loss. 

Consider a fixed context $x$ and define $\hat r(y|x)$ as:
\begin{equation}
\label{eq:r_hat}
\hat r(y|x) = \exp(r(x,y)),
\end{equation}
then our goal is to optimize ${\boldsymbol{\hat r}(\cdot|x)\in \mathbb R^\mathcal{I}_+}$, which is a $|\mathcal{I}|$-dim vector representing the latent rewards for all items in the recommendation dataset $\mathcal{I}$. we can rewrite the reward model's optimization as:
$$
\max_{\boldsymbol{\hat r}(\cdot|x)\in \mathbb R^\mathcal{I}_+} \sum_{y_w\in \mathcal{I}}\sum_{y_l\in \mathcal{I}}p_\mathcal{D}(y_w|x)q_\mathcal{D}(y_l|x)\log\left(\frac{\hat r(y_w|x)}{\hat r(y_w|x)+\hat r(y_l|x)}\right).
$$
Then we calculate the gradients:
\begin{align*}    
&\partial p_\mathcal{D}(y_w|x)q_\mathcal{D}(y_l|x)\log\left(\frac{\hat r(y_w|x)}{\hat r(y_w|x)+\hat r(y_l|x)}\right)/\partial \hat r(y|x)\\
=&\left\{
\begin{aligned}
& p_\mathcal{D}(y|x)q_\mathcal{D}(y_l|x)(\frac1{\hat r(y|x)} - \frac{1}{\hat r(y|x)+\hat r(y_l|x)}) && y_w=y,y_l\neq y,\\
& -p_\mathcal{D}(y_w|x)q_\mathcal{D}(y|x)\frac1{\hat r(y_w|x)+\hat r(y|x)} && y_w\neq y, y_l=y,\\
& 0 && \text{else.}
\end{aligned}
\right.
\end{align*}
Hence, the objective's gradient w.r.t. $\hat r(y|x)$ can be written as:
\begin{align*}
&\partial \sum_{y_w\in \mathcal{I}}\sum_{y_l\in \mathcal{I}}p_\mathcal{D}(y_w|x)q_\mathcal{D}(y_l|x)\log\left(\frac{\hat r(y_w|x)}{\hat r(y_w|x)+\hat r(y_l|x)}\right)/\partial \hat r(y|x)\\
=&0 + \sum_{y_l\neq y}\left[p_\mathcal{D}(y|x)q_\mathcal{D}(y_l|x)(\frac1{\hat r(y|x)} - \frac{1}{\hat r(y|x)+\hat r(y_l|x)})\right] \\
&\phantom{0+} - \sum_{y_w\neq y}p_\mathcal{D}(y_w|x)q_\mathcal{D}(y|x)\frac1{\hat r(y_w|x)+\hat r(y|x)}\\
=&\sum_{y_i\in \mathcal{I}}\left[p_\mathcal{D}(y|x)q_\mathcal{D}(y_i|x)(\frac1{\hat r(y|x)} - \frac{1}{\hat r(y|x)+\hat r(y_i|x)})\right] \\
&- \sum_{y_i\in \mathcal{I}}p_\mathcal{D}(y_i|x)q_\mathcal{D}(y|x)\frac1{\hat r(y_i|x)+\hat r(y|x)} \quad\text{(add 0)}\\
=&\sum_{y_i\in \mathcal{I}}\left[\frac{p_\mathcal{D}(y|x)q_\mathcal{D}(y_i|x)}{\hat r(y|x)} - \frac{p_\mathcal{D}(y_i|x)q_\mathcal{D}(y|x) + p_\mathcal{D}(y|x)q_\mathcal{D}(y_i|x)}{\hat r(y_i|x) + \hat r(y|x)}\right].
\end{align*}

By setting the gradients to be $0$, we obtain that for $\forall y \in \mathcal{I}$, the optimal reward $r^*(y|x)$ is:
$$
\hat r^*(y|x) \propto \frac{p_\mathcal{D}(y|x)}{q_\mathcal{D}(y|x)}.
$$

By plugging it into \myeq{r_hat}, we have:
$$
r^*(x,y) = \log p_\mathcal{D}(y|x) - \log q_\mathcal{D}(y|x)  + \text{Constant}.
$$

Back to the RLHF objective, we have:
\begin{gather*}
\max_{\theta} \mathbb E_{x\sim\mathcal D, y\sim\pi_\theta(\cdot|x)}[\log p_\mathcal{D}(y|x) - \log q_\mathcal{D}(y|x)] - \beta\mathrm{D_{KL}}(\pi_\theta\|\pi_\mathrm{ref}),
\end{gather*}
which has a well-known closed form solution \cite{rafailov2024direct}:
$$
\pi_\theta^*(y|x) \propto \pi_\mathrm{ref}(y|x)\cdot \left(\frac{p_\mathcal{D}(y|x)}{q_\mathcal{D}(y|x)}\right)^{1/\beta}.
$$
\end{proof}

\section{Dataset Statistics}
\label{appendix:datasets}

Our datasets span diverse domains, including movies, books, music, and games, offering varied sizes and user interaction patterns to provide a comprehensive basis for evaluating LRSs. Note that while we report the full dataset statistics, only a subset of interaction sequences is sampled for LLM fine-tuning, as detailed in \mysec{datasets}.

\begin{table}[h]
\centering
\caption{Statistics of Datasets.}
\label{tab:datasets}
\renewcommand{\arraystretch}{0.5}
\begin{tabular}{cccc}
\toprule
\multirow{2}{*}{\textbf{Datasets}} & \multirow{2}{*}{\textbf{\#Items}} & \multirow{2}{*}{\textbf{\#Interactions}} & \multirow{2}{*}{\textbf{\#Sequences}} \\
                                 &                         &                             &                            \\ \midrule
\multirow{2}{*}{\textbf{MovieLens}}       & \multirow{2}{*}{10,682} & \multirow{2}{*}{10,000,054} & \multirow{2}{*}{9,301,274} \\
                                 &                         &                             &                            \\
\multirow{2}{*}{\textbf{Goodreads}}       & \multirow{2}{*}{4,058}  & \multirow{2}{*}{160,398}    & \multirow{2}{*}{6,031}     \\
                                 &                         &                             &                            \\
\multirow{2}{*}{\textbf{CDs\_and\_Vinyl}} & \multirow{2}{*}{13,078} & \multirow{2}{*}{185,855}    & \multirow{2}{*}{21,347}    \\
                                 &                         &                             &                            \\
\multirow{2}{*}{\textbf{Steam}}           & \multirow{2}{*}{32,094} & \multirow{2}{*}{178,961}    & \multirow{2}{*}{29,876}    \\
                                 &                         &                             &                            \\ \bottomrule
\end{tabular}%
\end{table}

\section{Implementation Details}
\label{appendix:details}
For LLM-based methods, we adopted Llama-3.2-1B-Instruct as the backbone LLM. Considering the ability of LLMs to quickly adapt to downstream tasks with limited data, we followed BIGRec \cite{bao2023bi} and used relatively smaller datasets. To ensure fairness in comparison, all baseline methods and SPRec utilize the same dataset as used in the SFT training phase.
For SPRec, the total number of iterations was set to $5$, with each SFT and DPO phase trained for 1 epoch. To ensure that the training data used in each iteration is not identical, we further randomly sample half of the training data (i.e., 2048 interactions) for training in each iteration. All experiments were carried out on four RTX 3090 GPUs, each with 24GB of VRAM. 

For the traditional model SASRec, we use the same training and validation datasets as other LLM-based methods, with dataset sizes of 4,096 and 512, respectively. The embedding size was fixed at $64$, and the dropout ratio was set to 0.1. Negative samples were randomly sampled in training, with Adam as the optimizer and a learning rate of 4e-3.
More details of the implementation are available via \url{https://github.com/RegionCh/SPRec}.


\end{document}